%%%%%%%%%%%%%%%%%%%%%%%%%%%%%%%%%%%%%%%%%%
% determine hypertex mode
\newif\iflanl
\openin 1 lanlmac
\ifeof 1 \lanlfalse \else \lanltrue \fi
\closein 1
\iflanl
    \input lanlmac
\else
    \message{[lanlmac not found - use harvmac instead}
    \input harvmac
    \fi
\newif\ifhypertex
\ifx\hyperdef\UnDeFiNeD
    \hypertexfalse
    \message{[HYPERTEX MODE OFF}
    \def\hyperref#1#2#3#4{#4}
    \def\hyperdef#1#2#3#4{#4}
    \def\hypernoname{}
    \def\e@tf@ur#1{}
    \def\eprt#1{{\tt #1}}
    \def\CERN{\centerline{CERN, CH--1211 Geneva 23, Switzerland}}
    \def\wl{W.\ Lerche}
\else
    \hypertextrue
    \message{[HYPERTEX MODE ON}
%hypertex links to xxx.lanl.gov:
%  \def\hth/#1#2#3#4#5#6#7{\special{html:<a
%   href="http://xxx.lanl.gov/abs/hep-th/#1#2#3#4#5#6#7">}
%  {\tt hep-th/#1#2#3#4#5#6#7}\special{html:</a>}}
\def\eprt#1{{\tt
#1}}
\def\CERN{\centerline{

Theory Division, CERN, Geneva, Switzerland}}
\def\wl{
 W.\ Lerche}
\fi
%%%%%%%%%%%%%%%%%%%%%%% %%%%%%%%%%%%%%%%%%%%%%%
\newif\ifdraft

\noblackbox
\catcode`\@=11
\newif\iffrontpage
%%%%%%%%%%%%%%%%%%% %%%%%%%%%%%%%%%%%%%%%%%%%%%%%%%%%%%%%%%%%%%%%%
%%%%% sizes, offsets etc
%%%%%%%%%%%%%%%%%%% %%%%%%%%%%%%%%%%%%%%%%%%%%%%%%%%%%%%%%%%%%%%%%
\ifx\answ\bigans
\def\titleft{\titla}
\magnification=1200\baselineskip=14pt plus 2pt minus 1pt
%
%%%%% unreduced mode: %%%%
%\voffset=0.35truein\hoffset=0.250truein
\advance\hoffset by-0.075truein
\advance\voffset by1.truecm
\hsize=6.15truein\vsize=600.truept\hsbody=\hsize\hstitle=\hsize
\else\let\lr=L
\def\titleft{\titla}
\magnification=1000\baselineskip=14pt plus 2pt minus 1pt
%
%%%%% reduced mode: %%%%%%%
\hoffset=-0.75truein\voffset=-.0truein
%?\hoffset=-.25truein\voffset=-.0truein
\vsize=6.5truein
\hstitle=8.truein\hsbody=4.75truein
\fullhsize=10truein\hsize=\hsbody
\fi
\parskip=4pt plus 15pt minus 1pt
%
%%%%%%%%%%%%%%%%%%% %%%%%%%%%%%%%%%%%%%%%%%%%%%%%%%%%%%%%%%%%%%%%%
%%%%% figures
%%%%%%%%%%%%%%%%%%% %%%%%%%%%%%%%%%%%%%%%%%%%%%%%%%%%%%%%%%%%%%%%%
\newif\iffigureexists
\newif\ifepsfloaded
\def\epsfcheck{
\ifdraft% to speed up
\input epsf\epsfloadedtrue
\else
  \openin 1 epsf
  \ifeof 1 \epsfloadedfalse \else \epsfloadedtrue \fi
  \closein 1
  \ifepsfloaded
    \input epsf
  \else
\immediate\write20{NO EPSF FILE --- FIGURES WILL BE IGNORED}
  \fi
\fi
\def\epsfcheck{}}
\def\checkex#1{
\ifdraft
\figureexistsfalse\immediate%
\write20{Draftmode: figure #1 not included}
\figureexiststrue
\else\relax
    \ifepsfloaded \openin 1 #1
        \ifeof 1
           \figureexistsfalse
  \immediate\write20{FIGURE FILE #1 NOT FOUND}
        \else \figureexiststrue
        \fi \closein 1
    \else \figureexistsfalse
    \fi
\fi}
\def\missbox#1#2{$\vcenter{\hrule
\hbox{\vrule height#1\kern1.truein
\raise.5truein\hbox{#2} \kern1.truein \vrule} \hrule}$}
\def\lfig#1{%  this is to call the figure in the text
\let\labelflag=#1%
\def\numb@rone{#1}%
\ifx\labelflag\UnDeFiNeD%
{\xdef#1{\the\figno}%
\writedef{#1\leftbracket{\the\figno}}%
\global\advance\figno by1%
}\fi{\hyperref{}{figure}{{\numb@rone}}{Fig.{\numb@rone}}}}
\def\figinsert#1#2#3#4{%  this inserts the figure
\epsfcheck\checkex{#4}%
\def\figsize{#3}%
\let\flag=#1\ifx\flag\UnDeFiNeD
{\xdef#1{\the\figno}%
\writedef{#1\leftbracket{\the\figno}}%
\global\advance\figno by1%
}\fi
\goodbreak\midinsert%
\iffigureexists
\centerline{\epsfysize\figsize\epsfbox{#4}}%
\else%
\vskip.05truein
  \ifepsfloaded
  \ifdraft
  \centerline{\missbox\figsize{Draftmode: #4 not included}}%
  \else
  \centerline{\missbox\figsize{#4 not found}}
  \fi
  \else
  \centerline{\missbox\figsize{epsf.tex not found}}
  \fi
\vskip.05truein
\fi%
{\smallskip%
\leftskip 4pc \rightskip 4pc%
\noindent\ninepoint\sl \baselineskip=11pt%
{\bf{\hyperdef\hypernoname{figure}{{#1}}{Fig.{#1}}}:~}#2%
\smallskip}\bigskip\endinsert%
}

\def\boxit#1{\vbox{\hrule\hbox{\vrule\kern8pt
\vbox{\hbox{\kern8pt}\hbox{\vbox{#1}}\hbox{\kern8pt}}
\kern8pt\vrule}\hrule}}
\def\mathboxit#1{\vbox{\hrule\hbox{\vrule\kern8pt\vbox{\kern8pt
\hbox{$\displaystyle #1$}\kern8pt}\kern8pt\vrule}\hrule}}
%
%%%%%%%%%%%%%%%%%%% %%%%%%%%%%%%%%%%%%%%%%%%%%%%%%%%%%%%%%%%%%%%%%
%%%%%  fonts
%%%%%%%%%%%%%%%%%%% %%%%%%%%%%%%%%%%%%%%%%%%%%%%%%%%%%%%%%%%%%%%%%
\font\bigit=cmti10 scaled \magstep1

\font\titla=cmr10 scaled\magstep3
\font\tenmss=cmss10
\font\absmss=cmss10 scaled\magstep1

\newfam\mssfam
\font\footrm=cmr8  \font\footrms=cmr5
\font\footrmss=cmr5   \font\footi=cmmi8
\font\footis=cmmi5   \font\footiss=cmmi5
\font\footsy=cmsy8   \font\footsys=cmsy5
\font\footsyss=cmsy5   \font\footbf=cmbx8
\font\footmss=cmss8
\def\footfont{\def\rm{\fam0\footrm}
\textfont0=\footrm \scriptfont0=\footrms
\scriptscriptfont0=\footrmss
\textfont1=\footi \scriptfont1=\footis
\scriptscriptfont1=\footiss
\textfont2=\footsy \scriptfont2=\footsys
\scriptscriptfont2=\footsyss
\textfont\itfam=\footi \def\it{\fam\itfam\footi}
\textfont\mssfam=\footmss \def\mss{\fam\mssfam\footmss}
\textfont\bffam=\footbf \def\bf{\fam\bffam\footbf} \rm}
\def\tenpoint{\def\rm{\fam0\tenrm}
\textfont0=\tenrm \scriptfont0=\sevenrm
\scriptscriptfont0=\fiverm
\textfont1=\teni  \scriptfont1=\seveni
\scriptscriptfont1=\fivei
\textfont2=\tensy \scriptfont2=\sevensy
\scriptscriptfont2=\fivesy
\textfont\itfam=\tenit \def\it{\fam\itfam\tenit}
\textfont\mssfam=\tenmss \def\mss{\fam\mssfam\tenmss}
\textfont\bffam=\tenbf \def\bf{\fam\bffam\tenbf} \rm}
\ifx\answ\bigans\def\abstractfont{\tenpoint}\else
\def\abstractfont{\def\rm{\fam0\absrm}
\textfont0=\absrm \scriptfont0=\absrms
\scriptscriptfont0=\absrmss
\textfont1=\absi \scriptfont1=\absis
\scriptscriptfont1=\absiss
\textfont2=\abssy \scriptfont2=\abssys
\scriptscriptfont2=\abssyss
\textfont\itfam=\bigit \def\it{\fam\itfam\bigit}
\textfont\mssfam=\absmss \def\mss{\fam\mssfam\absmss}
\textfont\bffam=\absbf \def\bf{\fam\bffam\absbf}\rm}\fi
%
%%%%%%%%%%%%%%%%%%%%%%%%%%%%% %%%%%%%%%%%%%%%%%%%%%%%%%%%%%%%%%
%%%%% footnotes   (adapted from PHYZZX, no hypertext yet)
%%%%%%%%%%%%%%%%%%%%%%%%%%%%% %%%%%%%%%%%%%%%%%%%%%%%%%%%%%%%%%
\def\f@@t{\baselineskip10pt\lineskip0pt\lineskiplimit0pt
\bgroup\aftergroup\@foot\let\next}
\setbox\strutbox=\hbox{\vrule height 8.pt depth 3.5pt width\z@}
\def\vfootnote#1{\insert\footins\bgroup
\baselineskip10pt\footfont
\interlinepenalty=\interfootnotelinepenalty
\floatingpenalty=20000
\splittopskip=\ht\strutbox \boxmaxdepth=\dp\strutbox
\leftskip=24pt \rightskip=\z@skip
\parindent=12pt \parfillskip=0pt plus 1fil
\spaceskip=\z@skip \xspaceskip=\z@skip
\Textindent{$#1$}\footstrut\futurelet\next\fo@t}
\def\Textindent#1{\noindent\llap{#1\enspace}\ignorespaces}
\def\foot{\global\advance\ftno by1%
\attach{\hyperref{}{footnote}{\the\ftno}{\footsymbolgen}}%
\vfootnote{\hyperdef\hypernoname{footnote}{\the\ftno}{\footsymbol}}}%
%   this is for custom footnote marks:
\def\footnote#1{\global\advance\ftno by1%
\attach{\hyperref{}{footnote}{\the\ftno}{#1}}%
\vfootnote{\hyperdef\hypernoname{footnote}{\the\ftno}{#1}}}%
\newcount\lastf@@t           \lastf@@t=-1
\newcount\footsymbolcount    \footsymbolcount=0
\global\newcount\ftno \global\ftno=0
\def\footsymbolgen{\relax\footsym
\global\lastf@@t=\pageno\footsymbol}
\def\footsym{\ifnum\footsymbolcount<0
\global\footsymbolcount=0\fi
{\iffrontpage \else \advance\lastf@@t by 1 \fi
\ifnum\lastf@@t<\pageno \global\footsymbolcount=0
\else \global\advance\footsymbolcount by 1 \fi }
\ifcase\footsymbolcount
\fd@f\dagger\or \fd@f\diamond\or \fd@f\ddagger\or
\fd@f\natural\or \fd@f\ast\or \fd@f\bullet\or
\fd@f\star\or \fd@f\nabla\else \fd@f\dagger
\global\footsymbolcount=0 \fi }
\def\fd@f#1{\xdef\footsymbol{#1}}
\def\space@ver#1{\let\@sf=\empty \ifmmode #1\else \ifhmode
\edef\@sf{\spacefactor=\the\spacefactor}
\unskip${}#1$\relax\fi\fi}
\def\attach#1{\space@ver{\strut^{\mkern 2mu #1}}\@sf}
%
%%%%%%%%%%%%%%%%%%% %%%%%%%%%%%%%%%%%%%%%%%%%%%%%%%%%%%%%%%%%%%%%%
%%%%% References
%%%%%%%%%%%%%%%%%%% %%%%%%%%%%%%%%%%%%%%%%%%%%%%%%%%%%%%%%%%%%%%%%
\newif\ifnref
\def\rrr#1#2{\relax\ifnref\nref#1{#2}\else\ref#1{#2}\fi}
\def\ldf#1#2{\begingroup\obeylines
\gdef#1{\rrr{#1}{#2}}\endgroup\unskip}

\def\doubref#1#2{\refs{{#1},{#2}}}

\nreffalse
\def\refout{\listrefs}

\def\lref{\ldf}

%%%%%%%%%%%%%%%%%%% %%%%%%%%%%%%%%%%%%%%%%%%%%%%%%%%%%%%%%%%%%%%%%
%%%%%%% eq numbering
%%%%%%%%%%%%%%%%%%% %%%%%%%%%%%%%%%%%%%%%%%%%%%%%%%%%%%%%%%%%%%%%%
\def\eqn#1{\xdef #1{(\noexpand\hyperref{}%
{equation}{\secsym\the\meqno}%
{\secsym\the\meqno})}\eqno(\hyperdef\hypernoname{equation}%
{\secsym\the\meqno}{\secsym\the\meqno})\eqlabeL#1%
\writedef{#1\leftbracket#1}\global\advance\meqno by1}
\def\eqnalign#1{\xdef #1{\noexpand\hyperref{}{equation}%
{\secsym\the\meqno}{(\secsym\the\meqno)}}%
\writedef{#1\leftbracket#1}%
\hyperdef\hypernoname{equation}%
{\secsym\the\meqno}{\e@tf@ur#1}\eqlabeL{#1}%
\global\advance\meqno by1}
%old:
\def\eqnalign#1{\xdef #1{(\secsym\the\meqno)}
\writedef{#1\leftbracket#1}%
\global\advance\meqno by1 #1\eqlabeL{#1}}
%
%%%%%%%%%%%%%%%%%%% %%%%%%%%%%%%%%%%%%%%%%%%%%%%%%%%%%%%%%%%%%%%%%
%%%%%%  macros for titlepage, marginnotes, etc
%%%%%%%%%%%%%%%%%%% %%%%%%%%%%%%%%%%%%%%%%%%%%%%%%%%%%%%%%%%%%%%%%

%
\def\chap#1{\newsec{#1}}
\def\chapter#1{\chap{#1}}
\def\sect#1{\subsec{#1}}
\def\section#1{\sect{#1}}
\def\\{\ifnum\lastpenalty=-10000\relax
\else\hfil\penalty-10000\fi\ignorespaces}
\def\note#1{\leavevmode%
\edef\@@marginsf{\spacefactor=\the\spacefactor\relax}%
\ifdraft\strut\vadjust{%
\hbox to0pt{\hskip\hsize%
\ifx\answ\bigans\hskip.1in\else\hskip .1in\fi%
\vbox to0pt{\vskip-\dp
%\vskip4pt
\strutbox\sevenbf\baselineskip=8pt plus 1pt minus 1pt%
\ifx\answ\bigans\hsize=.7in\else\hsize=.35in\fi%
\tolerance=5000 \hbadness=5000%
\leftskip=0pt \rightskip=0pt \everypar={}%
\raggedright\parskip=0pt \parindent=0pt%
\vskip-\ht\strutbox\noindent\strut#1\par%
\vss}\hss}}\fi\@@marginsf\kern-.01cm}
\def\titlepage{%
\frontpagetrue\nopagenumbers\abstractfont%
\hsize=\hstitle\rightline{\vbox{\baselineskip=10pt%
{\abstractfont\pubnum}}}\pageno=0}
\frontpagefalse
\def\pubnum{}
\def\pdate{\number\month/\number\yearltd}
\def\makefootline{\iffrontpage\vskip .27truein
\line{\the\footline}
%\vskip -.1truein\line{\pdate\hfil}
\vskip -.1truein\leftline{\vbox{\baselineskip=10pt%
{\abstractfont\pdate}}}
\else\vskip.5cm\line{\hss \tenrm $-$ \folio\ $-$ \hss}\fi}
\def\title#1{\vskip .7truecm\titlestyle{\titleft #1}}
\def\titlestyle#1{\par\begingroup \interlinepenalty=9999
\leftskip=0.02\hsize plus 0.23\hsize minus 0.02\hsize
\rightskip=\leftskip \parfillskip=0pt
\hyphenpenalty=9000 \exhyphenpenalty=9000
\tolerance=9999 \pretolerance=9000
\spaceskip=0.333em \xspaceskip=0.5em
\noindent #1\par\endgroup }
\def\autskip{\ifx\answ\bigans\vskip.5truecm\else\vskip.1cm\fi}
\def\author#1{\vskip .7in \centerline{#1}}
\def\andauthor#1{\autskip
\centerline{\it and} \autskip\centerline{#1}}
\def\address#1{\ifx\answ\bigans\vskip.2truecm
\else\vskip.1cm\fi{\it \centerline{#1}}}
\def\abstract#1{
\vskip .5in\vfil\centerline
{\bf Abstract}\penalty1000
{{\smallskip\ifx\answ\bigans\leftskip 2pc \rightskip 2pc
\else\leftskip 5pc \rightskip 5pc\fi
\noindent\abstractfont \baselineskip=12pt
{#1} \smallskip}}
\penalty-1000}
\def\endpage{\tenpoint\supereject\global\hsize=\hsbody%
\frontpagefalse\footline={\hss\tenrm\folio\hss}}
\def\ack{\vskip2.cm\centerline{{\bf Acknowledgements}}}
%
%

%
%%%%%%%%%%%%%%%%%%%%%%%%%%%%% %%%%%%%%%%%%%%%%%%%%%%%%%%%%%%%%%
\def\bfone{\relax{\rm 1\kern-.35em 1}}
\def\inbar{\vrule height1.5ex width.4pt depth0pt}
\def\IC{\relax\,\hbox{$\inbar\kern-.3em{\mss C}$}}
\def\ID{\relax{\rm I\kern-.18em D}}
\def\IF{\relax{\rm I\kern-.18em F}}
\def\IH{\relax{\rm I\kern-.18em H}}
\def\II{\relax{\rm I\kern-.17em I}}
\def\IN{\relax{\rm I\kern-.18em N}}
\def\IP{\relax{\rm I\kern-.18em P}}
\def\IQ{\relax\,\hbox{$\inbar\kern-.3em{\rm Q}$}}
\def\IR{\relax{\rm I\kern-.18em R}}
\font\cmss=cmss10 \font\cmsss=cmss10 at 7pt
\def\ZZ{\relax\ifmmode\mathchoice
{\hbox{\cmss Z\kern-.4em Z}}{\hbox{\cmss Z\kern-.4em Z}}
{\lower.9pt\hbox{\cmsss Z\kern-.4em Z}}
{\lower1.2pt\hbox{\cmsss Z\kern-.4em Z}}\else{\cmss Z\kern-.4em
Z}\fi}
\def\a{\alpha}

\def\cC{{\cal C}} 
\def\cE{{\cal E}}

\def\cN{{\cal N}}

\def\nup#1({Nucl.\ Phys.\ $\us {B#1}$\ (}
\def\plt#1({Phys.\ Lett.\ $\us  {#1}$\ (}
\def\cmp#1({Comm.\ Math.\ Phys.\ $\us  {#1}$\ (}
\def\prp#1({Phys.\ Rep.\ $\us  {#1}$\ (}
\def\prl#1({Phys.\ Rev.\ Lett.\ $\us  {#1}$\ (}
\def\prv#1({Phys.\ Rev.\ $\us  {#1}$\ (}
\def\mpl#1({Mod.\ Phys.\ Let.\ $\us  {A#1}$\ (}
\def\ijmp#1({Int.\ J.\ Mod.\ Phys.\ $\us{A#1}$\ (}
\def\tit#1|{{\it #1},\ }
%
%%%%%%%%%%%%%%%%%%%%%%%%%%%%%%%% %%%%%%%%%%%%%%%%%%%%%%%%%%%%%%
%%%%% misc %%%%
%%%%%%%%%%%%%%%%%%%%%%%%%%%%%%%% %%%%%%%%%%%%%%%%%%%%%%%%%%%%%%

%

\def\ni{\noindent}

\def\bar{\overline}
\def\us#1{\underline{#1}}

\def\hat{\widehat}

\def\Coe#1.#2.{{#1\over #2}}

\def\coe#1.#2.{\relax{\textstyle {#1 \over #2}}\displaystyle}

\def\to{\rightarrow}
\def\notin{\hbox{{$\in$}\kern-.51em\hbox{/}}}

\def\del{\partial}

%%%%%%%%%%%%%%%%%%%%%%%%%%%

%%%%%%%%%%%%%%%%%%%%%%%%%%%
\catcode`\@=12
%%%%%%%%% end macros  %%%%%%% %%%%%%%%%%%%%%%%%%%%%%%%%%%%%%
%%%%%%%%%%%%%%%%%%%%%%%%%%%% %%%%%%%%%%%%%%%%%%%%%%%%%%%%%

%%%%%%%%%%%%%%%%%%%%%%%%%%%% %%%%%%%%%%%%%%%%%%%%%%%%%%%%%
% references: 
%%%%%%%%%%%%%%%%%%%%%%%%%%%% %%%%%%%%%%%%%%%%%%%%%%%%%%%%%

%\def\eprt#1{{\tt #1}}
\def\nihil#1{{\sl #1}}
\def\br{\hfill\break}

\def\ijmp {{Int. J. Mod. Phys.\ }{\bf A}}

\lref\arn{V.\ Arnold, S.\ Gusein-Zade and A.\ Varchenko,
{\it Singularities of Differentiable Maps},
Birkh\"auser, 1988.}

\lref\DF{M.\ R.\ Douglas and B.\ Fiol, 
\nihil{D-branes and discrete torsion.\ II,}
\eprt{hep-th/9903031}. 
%%CITATION = HEP-TH 9903031;%%
}

\lref\BDLR{I.\ Brunner, M.R.\ Douglas, A.\ Lawrence and 
C.\ R\"omelsberger, 
\nihil{D-branes on the quintic,}
\eprt{hep-th/9906200}. 
%%CITATION = HEP-TH 9906200;%%
}

\lref\OOY{H.\ Ooguri, Y.\ Oz and Z.\ Yin, 
\nihil{D-branes on Calabi-Yau spaces and their mirrors,}
 Nucl.\ Phys.\ {\bf B477} 407 (1996), 
\eprt{hep-th/9606112}. 
%%CITATION = HEP-TH 9606112;%%
}

\lref\bound{
See e.g.,:
{A.\ Recknagel and V.\ Schomerus, 
\nihil{D-branes in Gepner models,}
 Nucl.\ Phys.\ {\bf B531} 185 (1998), 
\eprt{hep-th/9712186};\br 
%%CITATION = NUPHA,B531,185;%%
}
{J.\ Fuchs and C.\ Schweigert, 
\nihil{Branes: From free fields to general backgrounds,}
 Nucl.\ Phys.\ {\bf B530} 99 (1998), 
\eprt{hep-th/9712257}. 
%%CITATION = HEP-TH 9712257;%%
}
}

\lref\DALE{
{C.\ M.\ Hull and P.\ K.\ Townsend, 
\nihil{Unity of superstring dualities,}
 Nucl.\ Phys.\ {\bf B438} 109 (1995), 
\eprt{hep-th/9410167}; 
%%CITATION = HEP-TH 9410167;%%
}
{%C.\ M.\ Hull and P.\ K.\ Townsend, 
\nihil{Enhanced gauge symmetries in superstring theories,}
 Nucl.\ Phys.\ {\bf B451} 525 (1995), 
\eprt{hep-th/9505073}; 
%%CITATION = HEP-TH 9505073;%%
}\br
{E.\ Witten, 
\nihil{Some comments on string dynamics,}
Contributed to STRINGS 95: 
{\it Future Perspectives in String Theory}, 
Los Angeles, CA, 13-18 Mar 1995,
\eprt{hep-th/9507121}. 
%%CITATION = HEP-TH 9507121;%%
}
}

\lref\gravdesc{
{K.\ Li, 
\nihil{Topological gravity with minimal matter,}
 Nucl.\ Phys.\ {\bf B354} (1991) 711; 
%%CITATION = NUPHA,B354,711;%%
}\br
{S.\ Govindarajan, T.\ Jayaraman and V.\ John, 
\nihil{Chiral rings and physical states in $c<1$ string theory,}
Nucl.\ Phys.\ {\bf B402} 118 (1993), 
\eprt{hep-th/9207109}; 
%%CITATION = HEP-TH 9207109.%%
}
}

\lref\equivc{
{A. Losev, 
\nihil{Descendants constructed from matter field and 
K.Saito higher residue pairing in Landau-Ginzburg 
theories coupled to topological gravity},
preprint TPI-MINN-92-40-T;
}\br
{M.\ Bershadsky, W.\ Lerche, D.\ Nemeschansky and N.\ P.\ Warner, 
\nihil{Extended N=2 superconformal structure of 
gravity and W gravity coupled to matter,}
 Nucl.\ Phys.\ {\bf B401} 304 (1993), 
\eprt{hep-th/9211040}. 
%%CITATION = HEP-TH 9211040;%%
}\br
{T.\ Eguchi, H.\ Kanno, Y.\ Yamada and S.\ Yang, 
\nihil{Topological strings, flat coordinates and 
gravitational descendants,}
 Phys.\ Lett.\ {\bf B305} 235 (1993), 
\eprt{hep-th/9302048}. 
%%CITATION = HEP-TH 9302048;%%
}
}

\lref\ground{E.\ Witten, 
\nihil{Ground ring of two-dimensional string theory,}
 Nucl.\ Phys.\ {\bf B373} 187 (1992), 
\eprt{hep-th/9108004}. 
%%CITATION = HEP-TH 9108004;%%
}

\lref\geomeng{
{S.\ Katz, A.\ Klemm and C.\ Vafa, 
\nihil{Geometric engineering of quantum field theories,}
 Nucl.\ Phys.\ {\bf B497} 173 (1997), 
\eprt{hep-th/9609239};\br 
%%CITATION = HEP-TH 9609239;%%
}
{S.\ Katz, P.\ Mayr and C.\ Vafa, 
\nihil{Mirror symmetry and exact solution of 4D N = 2 gauge theories.\ I,}
 Adv.\ Theor.\ Math.\ Phys.\ {\bf 1} 53 (1998), 
\eprt{hep-th/9706110}. 
%%CITATION = HEP-TH 9706110;%%
}
}

\lref\BLG{
{K.\ Hori and C.\ Vafa, 
\nihil{Mirror symmetry,}
\eprt{hep-th/0002222}; 
%%CITATION = HEP-TH 0002222;%%
}
\br
{S.\ Govindarajan and T.\ Jayaraman, 
\nihil{On the Landau-Ginzburg description of boundary CFTs and special 
Lagrangian submanifolds,}
\eprt{hep-th/0003242}; 
%%CITATION = HEP-TH 0003242;%%
}
{K.\ Hori, A.\ Iqbal and C.\ Vafa, 
\nihil{D-branes and mirror symmetry,}
\eprt{hep-th/0005247}. 
%%CITATION = HEP-TH 0005247;%%
}
}

\lref\HIV{K.\ Hori, A.\ Iqbal and C.\ Vafa, as cited in \BLG.}

\lref\SS{
{C.\ Vafa and N.\ Warner, 
\nihil{Catastrophes And The Classification Of Conformal Theories,}
 Phys.\ Lett.\ {\bf B218} 51 (1989);\br 
%%CITATION = PHLTA,B218,51;%%
}
{E.\ J.\ Martinec, 
\nihil{Criticality, Catastrophes And Compactifications,}
 Print-89-0373 (EFI,CHICAGO), {\sl  In Brink, L.\ (ed.) et al.: 
Physics and mathematics of strings 389-433.}
}
}

\lref\MNMN{M.\ Naka and M.\ Nozaki, 
\nihil{Boundary states in Gepner models,}
 JHEP{\bf 0005} 027 (2000), 
\eprt{hep-th/0001037}. 
%%CITATION = HEP-TH 0001037;%%
}

\lref\GK{A.\ Giveon and D.\ Kutasov, 
\nihil{Little string theory in a double scaling limit,}
 JHEP{\bf 9910} 034 (1999), 
\eprt{hep-th/9909110}. 
%%CITATION = HEP-TH 9909110;%%
}

\lref\OV{H.\ Ooguri and C.\ Vafa, 
\nihil{Two-Dimensional Black Hole and Singularities of CY Manifolds,}
 Nucl.\ Phys.\ {\bf B463} 55 (1996), 
\eprt{hep-th/9511164}. 
%%CITATION = HEP-TH 9511164;%%
}

\lref\OOY{H.\ Ooguri, Y.\ Oz and Z.\ Yin, 
\nihil{D-branes on Calabi-Yau spaces and their mirrors,}
 Nucl.\ Phys.\ {\bf B477} 407 (1996), 
\eprt{hep-th/9606112}. 
%%CITATION = HEP-TH 9606112;%%
}

\lref\KLM{A.\ Klemm, W.\ Lerche and P.\ Mayr, 
\nihil{K3 Fibrations and heterotic type II string duality,}
 Phys.\ Lett.\ {\bf B357} 313 (1995), 
\eprt{hep-th/9506112}. 
%%CITATION = HEP-TH 9506112;%%
}

\lref\KLMVW{A.\ Klemm, W.\ Lerche, P.\ Mayr, C.\ Vafa and N.\ Warner, 
\nihil{Self-Dual Strings and N=2 Supersymmetric Field Theory,}
 Nucl.\ Phys.\ {\bf B477} 746 (1996), 
\eprt{hep-th/9604034}. 
%%CITATION = HEP-TH 9604034;%%
}

\lref\curves{
{A.\ Klemm, W.\ Lerche, S.\ Yankielowicz and S.\ Theisen, 
\nihil{Simple singularities and N=2 supersymmetric Yang-Mills theory,}
 Phys.\ Lett.\ {\bf B344} 169 (1995), 
\eprt{hep-th/9411048};\br
%%CITATION = HEP-TH 9411048;%%
}
{P.\ C.\ Argyres and A.\ E.\ Faraggi, 
\nihil{The vacuum structure and spectrum of N=2 
supersymmetric SU(n) gauge theory,}
 Phys.\ Rev.\ Lett.\ {\bf 74} 3931 (1995), 
\eprt{hep-th/9411057};\br 
%%CITATION = HEP-TH 9411057;%%
}
{E.\ Martinec and N.\ Warner, 
\nihil{Integrable systems and supersymmetric gauge theory,}
 Nucl.\ Phys.\ {\bf B459} 97 (1996), 
\eprt{hep-th/9509161}. 
%%CITATION = HEP-TH 9509161;%%
}
}

\lref\zub{J.\ Zuber, 
\nihil{CFT, BCFT, ADE and all that,}
\eprt{hep-th/0006151}. 
%%CITATION = HEP-TH 0006151;%%
}

\lref\BPPZ{R.\ E.\ Behrend, P.\ A.\ Pearce, V.\ B.\ Petkova and J.\ Zuber, 
\nihil{Boundary conditions in rational conformal field theories,}
 Nucl.\ Phys.\ {\bf B570} 525 (2000), 
\eprt{hep-th/9908036}. 
%%CITATION = HEP-TH 9908036;%%
}

\lref\CIZ{
{A.\ Cappelli, C.\ Itzykson and J.\ B.\ Zuber, 
\nihil{Modular Invariant Partition Functions In Two-Dimensions,}
 Nucl.\ Phys.\ {\bf B280} 445 (1987); 
%%CITATION = NUPHA,B280,445;%%
}
{% A.\ Cappelli, C.\ Itzykson and J.\ B.\ Zuber, 
\nihil{The ADE Classification Of Minimal And $A_1^{(1)}$ 
 Conformal Invariant Theories,}
 Commun.\ Math.\ Phys.\ {\bf 113} 1 (1987). 
%%CITATION = CMPHA,113,1;%%
}
}

\lref\Kos{B.\ Kostant, 
\nihil{The principal three-dimensional subgroup and the
Betti numbers of a complex Lie group},
Am.\ J.\ Math.\ 81 (1959) 973.}

\lref\hump{
H.\ Hiller, \nihil{Geometry of Coxeter groups,}
Pitman London 1982; \br
J.\ Humphreys, 
\nihil{Reflection groups and Coxeter groups,}
Cambridge University Press 1990.}

\lref\dorAB{
{P.\ Dorey, 
\nihil{Root systems and purely elastic S matrices,}
 Nucl.\ Phys.\ {\bf B358} 654 (1991); 
%%CITATION = NUPHA,B358,654;%%
}
{%P.\ Dorey, 
\nihil{Root systems and purely elastic S matrices.\ 2,}
 Nucl.\ Phys.\ {\bf B374} 741 (1992), 
\eprt{hep-th/9110058}. 
%%CITATION = HEP-TH 9110058;%%
}
}

\lref\dorC{P.\ Dorey, 
\nihil{Partition functions, intertwiners and the Coxeter element,}
 Int.\ J.\ Mod.\ Phys.\ {\bf A8} 193 (1993), 
\eprt{hep-th/9205040}. 
%%CITATION = HEP-TH 9205040;%%
}

\lref\LW{W.\ Lerche and N.\ P.\ Warner, 
\nihil{Polytopes and solitons in integrable, 
N=2 supersymmetric Landau-Ginzburg theories,}
 Nucl.\ Phys.\ {\bf B358} 571 (1991). 
%%CITATION = NUPHA,B358,571;%%
}

\lref\WL{W.\ Lerche, 
\nihil{On a boundary CFT description of 
nonperturbative N = 2 Yang-Mills theory,}
\eprt{hep-th/0006100}. 
%%CITATION = HEP-TH 0006100;%%
}

\lref\cardy{J.\ L.\ Cardy, 
\nihil{Boundary Conditions, Fusion Rules And The Verlinde Formula,}
 Nucl.\ Phys.\ {\bf B324} 581 (1989). 
%%CITATION = NUPHA,B324,581;%%
}

\lref\otherALE{
{D.\ Anselmi, M.\ Billo, P.\ Fre, L.\ Girardello and A.\ Zaffaroni, 
\nihil{ALE manifolds and conformal field theories,}
 Int.\ J.\ Mod.\ Phys.\ {\bf A9} 3007 (1994), 
\eprt{hep-th/9304135}; 
%%CITATION = HEP-TH 9304135;%%
}\br
{M.\ Bershadsky, C.\ Vafa and V.\ Sadov, 
\nihil{D-Strings on D-Manifolds,}
 Nucl.\ Phys.\ {\bf B463} 398 (1996), 
\eprt{hep-th/9510225}; 
%%CITATION = HEP-TH 9510225;%%
}\br
{H.\ Ooguri and C.\ Vafa, 
\nihil{Two-Dimensional Black Hole and Singularities of CY Manifolds,}
 Nucl.\ Phys.\ {\bf B463} 55 (1996), 
\eprt{hep-th/9511164}; 
%%CITATION = HEP-TH 9511164;%%
}\br
{C.\ Vafa, 
\nihil{Instantons on D-branes,}
 Nucl.\ Phys.\ {\bf B463} 435 (1996), 
\eprt{hep-th/9512078}; 
%%CITATION = HEP-TH 9512078;%%
}\br
{C.\ V.\ Johnson and R.\ C.\ Myers, 
\nihil{Aspects of type IIB theory on ALE spaces,}
 Phys.\ Rev.\ {\bf D55} 6382 (1997), 
\eprt{hep-th/9610140}; 
%%CITATION = HEP-TH 9610140;%%
}\br
{M.\ R.\ Douglas and G.\ Moore, 
\nihil{D-branes, Quivers, and ALE Instantons,}
\eprt{hep-th/9603167}; 
%%CITATION = HEP-TH 9603167;%%
}\br
{J.\ A.\ Harvey and G.\ Moore, 
\nihil{On the algebras of BPS states,}
 Commun.\ Math.\ Phys.\ {\bf 197} 489 (1998), 
\eprt{hep-th/9609017}; 
%%CITATION = HEP-TH 9609017;%%
}\br
{M.\ R.\ Douglas, 
\nihil{Enhanced gauge symmetry in M(atrix) theory,}
 JHEP{\bf 9707} 004 (1997), 
\eprt{hep-th/9612126}; 
%%CITATION = HEP-TH 9612126;%%
}\br
{D.\ Diaconescu and J.\ Gomis, 
\nihil{Duality in matrix theory and three 
dimensional mirror symmetry,}
 Nucl.\ Phys.\ {\bf B517} 53 (1998), 
\eprt{hep-th/9707019}; 
%%CITATION = HEP-TH 9707019;%%
}
\br
{T.\ Takayanagi, 
\nihil{String creation and monodromy from 
fractional D-branes on ALE spaces,}
 JHEP{\bf 0002} 040 (2000), 
\eprt{hep-th/9912157}; 
%%CITATION = HEP-TH 9912157;%%
}\br
{B.\ Fiol and M.\ Marino, 
\nihil{BPS states and algebras from quivers,}
\eprt{hep-th/0006189}. 
%%CITATION = HEP-TH 0006189;%%
}
}

\lref\naka{H. Nakajima, 
\nihil{Instantons on ALE spaces, quiver varieties, 
and Kac-Moody algebras,}
Duke Math.~{\bf 76} (1994) 365;
\nihil{Gauge theory on resolutions of simple singularities
and affine Lie algebras,}
Int.\ Math.\ Res.\ Not.\ {\bf 2} (1994) 61;
\nihil{Instantons and affine Lie algebras,}
\eprt{alg-geom/9502013.}
}

\lref\FM{B.\ Fiol and M.\ Marino, 
as cited in \otherALE.}

\lref\DR{D.\ Diaconescu and C.\ R\"omelsberger, 
\nihil{D-branes and bundles on elliptic fibrations,}
\eprt{hep-th/9910172}.
%%CITATION = HEP-TH 9910172;%%
}

\lref\Taka{T.\ Takayanagi, as in \otherALE.}

\lref\bcond{
{J.\ Fuchs and C.\ Schweigert, 
\nihil{Solitonic sectors, alpha-induction 
and symmetry breaking boundaries,}
\eprt{hep-th/0006181}. 
%%CITATION = HEP-TH 0006181;%%
}
}

\lref\gannon{T.\ Gannon, 
\nihil{$U(1)^m$ modular invariants, $N = 2$ minimal models, 
and the quantum Hall effect,}
 Nucl.\ Phys.\ {\bf B491} 659 (1997), 
\eprt{hep-th/9608063}. 
%%CITATION = HEP-TH 9608063;%%
}

\lref\eguchi{
{T.\ Eguchi and Y.\ Sugawara, 
\nihil{Modular invariance in superstring 
on Calabi-Yau n-fold with A-D-E singularity,}
 Nucl.\ Phys.\ {\bf B577} 3 (2000), 
\eprt{hep-th/0002100}; 
%%CITATION = HEP-TH 0002100;%%
}
\br
{S.\ Mizoguchi, 
\nihil{Modular invariant critical superstrings 
on four-dimensional Minkowski space x two-dimensional black hole,}
 JHEP{\bf 0004} 014 (2000), 
\eprt{hep-th/0003053}. 
%%CITATION = HEP-TH 0003053;%%
}
}

\lref\pelc{O.\ Pelc, 
\nihil{Holography, singularities on orbifolds and 4D N = 2 SQCD,}
 JHEP{\bf 0003} 012 (2000), 
\eprt{hep-th/0001054}. 
%%CITATION = HEP-TH 0001054;%%
}

%%%%%%% paper  specific macros

\def\N{{\cN}}
\def\IG{\relax\,\hbox{$\inbar\kern-.3em{\mss G}$}}
\def\ep{\eta}
\def\cW{{\cal W}}

%%%%%%%%%%%%%%%%%%%%%%%%%%%% %%%%%%%%%%%%%%%%%%%%%%%%%%%%%

%\draft

\def\pubnum{
\hbox{CERN-TH/2000-178}
\hbox{hep-th/0006247}
\hbox{}}
\def\pdate{}
\titlepage
\vskip2.cm
\title{{\titlefont $D$-Branes on ALE Spaces
and the ADE Classification
of Conformal Field Theories}}
\vskip -.7cm
\autskip
\author{\wl, {C.A.\ L\"utken}\foot{
On leave from Dept.\ of Physics, 
University of Oslo, N-0316 Oslo, Norway}}
 \vskip.2truecm
\CERN
\andauthor{C.\ Schweigert}
 \vskip.2truecm
\centerline{LPTHE, Universit\'e Paris VI, 
4 place Jussieu, F-75252 Paris, Cedex 05, France}

\vskip-.2truecm

\abstract{
The spectrum of $D2$-branes wrapped on an ALE space of
general ADE type is determined, by representing them as 
boundary states of $\N\!=\!2$ superconformal minimal models. 
The stable quantum states have RR charges which precisely
represent the gauge fields of the corresponding Lie algebra. 
This provides a simple and direct physical link between the 
ADE classification of $\N\!=\!2$ superconformal field theories, 
and the corresponding root systems. An affine extension of this 
structure is also considered, whose boundary states
represent the $D2$-branes plus additional $D0$-branes.
}

\vfil
%\vskip 1.cm
\ni {CERN-TH/2000-178}\hfill\break
\ni June 2000
\endpage
\baselineskip=14pt plus 2pt minus 1pt

\sequentialequations

%%%%%%%%%%%%%%%%%%%%%%%%%%%%%%%%%%%%%%%%%%%%%
\chapter{Introduction}
%%%%%%%%%%%%%%%%%%%%%%%%%%%%%%%%%%%%%%%%%%%%%

After their r\^ole in ``minimal'' conformal field theories had been
discovered~\CIZ, ADE\foot{As usual, this stands for the simply laced
Lie algebras of type $A_n$, $D_n$, $E_{6,7,8}$.} classifications have
surfaced in various manifestations in the physics literature; for a
recent review, see \zub. However, what has been lacking for a long time is
a deeper insight into how these various manifestations are related to
each other. In particular, while the CFT partition functions encode
some group theoretical data (most notably the exponents $\ep_j$, 
which label the diagonal terms), 
more basic group theoretical quantities like root
systems could not be identified within the (bulk) conformal field
theories. 
It is, however, an old observation \dorC\ that other basic data, in
the form of eigenvectors of the Cartan matrix, play a r\^ole in
constructing boundary states associated with the conformal field
theories. This point has been stressed recently again in \BPPZ.

That root systems of ADE type appear indeed more naturally on the
boundary rather than in the bulk, can easily be seen as follows. As is
well known, the ADE classification of modular invariants of $SU(2)_k$
can be directly related to modular invariants of the $\N=2$
superconformal minimal models by representing these in terms of cosets\foot
{Note that one can construct further partition functions
from this coset which do not describe the $\N=2$ minimal models \gannon.} 
$(SU(2)_k\!\times\! U(1))/U(1)$). The latter can in turn be
formulated \SS\ as Landau-Ginzburg models based on superpotentials
given by Arnold's simple singularities of ADE type \arn. Geometrically,
the middle-dimensional homology of the resolution of these
singularities is known to be isomorphic to the corresponding root
lattices. More precisely:  $H_*(M,\ZZ)\cong\Gamma_R^{(ADE)}$,
where $M\equiv M^{(ADE)}$ denotes an ALE space which is a non-compact model of
an ADE singularity on a compact $K3$ manifold. 

Physically this means that $D2$-branes wrapped around the vanishing
cycles of $M$ carry the RR quantum numbers of the charged
gauge fields of the corresponding ADE type \DALE. A priori, the wrapped
$D$-branes are solitonic objects on which open strings end, but they
can be represented as boundary states of the CFT.  The boundary
states thus probe the homology $H_*(M,\ZZ)$ of the ALE space 
and so exhibit somewhat finer details of the geometry
than the bulk chiral ring, which probes $H^*_{\bar\del}(M,\IC)$. This
then completes the chain of connections between the ADE classification
of modular invariants and  $D$-brane configurations of ADE type, i.e.,
root systems. \foot{From this point of view, the fact that the
exponents $\ep_j$ of the Lie algebra appear in  the bulk partition
function is simply a consequence  of the fact that the Ishibashi states
that underlie the boundary states are determined by the diagonal terms
of the bulk partition function.}

Using boundary Landau-Ginzburg theory \BLG,
this line of thoughts has been used recently \WL\ to analyze
the spectrum of quantum $D$-branes on ALE spaces and Seiberg-Witten
curves. However, the discussion there was restricted to
gauge symmetries of type $SU(N)\sim A_{N-1}$. The purpose
of the present brief note is to extend the computation
in a systematic and uniform way to all the simply laced Lie
algebras. 

A convenient setting of the problem is to consider type IIA strings
compactified on ALE spaces 
(for related works on strings and $D$-branes on
ALE spaces see e.g., \otherALE). Such string compactifications can be
described in conformal field theory by Landau-Ginzburg superpotentials
of the form \OV: $W_{G}(z,x_i,u_k)={ z^{-h}}+P_{G}(x_1,x_2,x_3,u_k)$, where
$h=h(G)$ and $P_{G}(\cdot,u_k)$ are the dual Coxeter number and simple
singularity, respectively, of the corresponding simply laced Lie
algebra $G$ of type ADE. More specifically, we will be here interested
in the exactly solvable ``Gepner'' points in the respective moduli
spaces, described by the following superpotentials \OV:
$$
\eqalign{W_{A_{h-1}}\ &=\  z^{-h} + {x_1}^h + {x_2}^2 + {x_3}^2  \cr
         W_{D_{{h\over 2}+1}}\  &=\  z^{-h} 
         +{x_1}^{{h\over2}}+ x_1 {x_2}^2 + {x_3}^2  \cr
         W_{E_6}\  &=\  z^{-12} + {x_1}^4 +{x_2}^3 + {x_3}^2  \cr
         W_{E_7}\  &=\  z^{-18} + {x_1}^3 {x_2} +x_2^3 + {x_3}^2  \cr
         W_{E_8}\  &=\  z^{-30} + {x_1}^5 +{x_2}^3 + {x_3}^2\ .
}\eqn\LGpotentials
$$
These non-compact Landau-Ginzburg theories describe smooth conformal
field theories with $\hat c=2$, which can also be represented in terms
of coset models based on $({SU(2)_{h-2}\over U(1)}\times
{SL(2)_{h+2}\over U(1)})/\ZZ_h$ \refs{\OV,\GK,\eguchi}. The non-compact
$z$-dependent piece, corresponding to the  $SL(2)$ factor in the coset,
describes the non-universal degrees of freedom that are not important
for our purposes; its r\^ole is mainly to push the central charge up to
the right value, and also to supply a certain contribution to the
intersection form that we are going to compute in the present letter.

%%%%%%%%%%%%%%%%%%%%%%%%%%%%%%%%%%%%%%%%%%%%%%%%%%%%%%%
\chapter{Boundary state intersection index}
%%%%%%%%%%%%%%%%%%%%%%%%%%%%%%%%%%%%%%%%%%%%%%%%%%%%%%%

We will now compute the topological intersection index
\doubref\DF\BDLR\  $I_{a,b}\equiv {\rm Tr}_{a,b}[(-1)^F]$ of the
boundary states associated with the LG theories defined in
\LGpotentials. Since these theories are tensor products of the $\N=2$
minimal models and their non-compact counterparts, we begin by
discussing the boundary states 
\bound\ of the minimal models based on $SU(2)_{k}\over
U(1)$, at levels $k=h-2$. They are labelled by $|L,M,S\rangle$, where
$L=1,...,r\equiv {\rm rank}(G)$, $M=-h+1,...,h$ (mod $2h$),  and
finally $S=-1,0,1,2$ (mod $4$) determines the R- or NS-sectors
(there is a selection rule that puts a constraint
on the labels $L,M,S$, whose precise form
will be discussed later). The boundary states can be expanded into the
Ishibashi states $|\ \rangle\rangle$ as follows \MNMN:
$$
\big|L,M,S\big\rangle\ =\ \sum_{(\ell,m,s)}
{
\psi_{L}^{\ (\ell)}
\over
\sqrt{{S_{(0,0,0)}^{\ \ (\ell,m,s)}}}
}\,e^{i{\pi\over h}(mM-{h\over 2}sS)}
\big|\ell,m,s \big\rangle\big\rangle\ 
\eqn\Ishi
$$
(up to normalization). Here, the Ishibashi labels $\ell$ run over the
exponents\foot{Explicitly, exponents $\ep_j$ and Coxeter numbers $h$ are: 
$\cE(A_n) = \{1, 2, \ldots , n\}$, $h(A_n)=n+1$,
$\cE(D_n) = \{1, 3, 5, \ldots , 2n-3,n-1 \}$, 
$h(D_n)=2n-2$,
$\cE(E_6) = \{1, 4, 5,7, 8, 11 \}$, $h(E_6)=12$,
$\cE(E_7) = \{1, 5, 7, 9, 11, 13, 17\}$, $h(E_7)=18$, and 
$\cE(E_8) = \{ 1, 7, 11, 13, 17,19, 23, 29 \}$, $h(E_8)=30$,
respectively.}  
$\ep_j$ associated with the Lie algebra $G$:
$(\ell+1)\in\{\ep_j\}\equiv \cE(G)$  ($m$ and $s$ run like $M$ and $S$ as
explained above), and
$$
S_{(\ell,m,s)}^{\ \ (\ell',m',s')}={1\over\sqrt2h}
\sin\big[{\pi\over h}(\ell+1)(\ell'+1)\big]
\exp\big[{i{\pi\over h}(mm'-{h\over2}ss')}\big]
\eqn\Smat
$$
are the modular transformation matrices of the $\N=2$ characters.
Moreover, $\psi_{L}^{\ (\ell)}$ are the orthonormal
eigenvectors of the ADE Cartan matrix $\cC(G)$, with eigenvalues
$\gamma_{\ell}\equiv 2-2\cos[\pi\ep_{\ell}/h]$, $\ell=1,...,r$
It was shown in \BPPZ\ that the analogous boundary states $|L\rangle$
of $\widehat{SU(2)}_{k=h-2}$ indeed solve the Cardy condition~\cardy.

Note that the labels $\ell$ and $L$ are in general on logically
different footings: while $(\ell+1)$ labels the exponents $\ep_j(G)$,
$L$ labels the components of the eigenvectors of $\cC(G)$.  
This is in accordance
with what was said in the introduction: namely that the boundary states
are naturally associated with the homology lattice, $H_*(M,\ZZ)$, which
in the present situation is given by the root lattice. On the other
hand, the bulk chiral ring, which is isomorphic to the cohomology ring
$H^*_{\bar\del}(M,\IC)$, is associated with
the exponents of the simple singularity. Only for the $A$ series there
is no distinction between the bulk and the boundary fusion algebras,
that is, between the indices $L$ and $\ell$ and between $\psi_{L}^{\
(\ell')}$ and $S_{\ell}^{\ (\ell')}$.

We will now adopt a particular convention for labeling the root
system, following \hump. That is, we split the simple roots $\a_i$
according to a  bi-coloration of the Dynkin diagram, so that we obtain
two orthogonal subsets of mutually commuting roots.\foot{
If $G$ is of type $D_{{\rm even}}$ or $E_{6,8}$, these two sets
correspond to symmetry preserving and symmetry breaking
boundary conditions, as discussed in \bcond.} Following the
notation of ref.\ \dorC, we  represent this in the following way:
$\{\a_i\}\equiv\{a_\bullet\}\cup\{a_\circ\}$. Given these two sets of
labels, we resolve the sign ambiguities of the eigenvectors as follows:
$$
\psi_{\bullet}^{\ (\ell)}\ =\ \psi_{\bullet}^{\ (h-\ell)}\ ,
\qquad
\psi_{\circ}^{\ (\ell)}\ =-\psi_{\circ}^{\ (h-\ell)}\ .
\eqn\signchoice
$$

The intersection index can now be
computed \doubref\BDLR\HIV\ by evaluating an 
overlap amplitude in the $RR$-sector: 
$$
I(L_1,L_2,M_1,M_2,S_1,S_2)\ \equiv\
{\phantom{\big|}}_{{\rm RR}}\big\langle L_1,M_1,S_1\big|
L_2,M_2,S_2\big\rangle_{{\rm RR}}\ .
\eqn\interdef
$$
For fixed $S$ and $L$ labels, this can be viewed as a $2h\times 2h$
matrix whose components are labeled by the $M_i$. Inserting the
expansion \Ishi, restricting to the $RR$ ground states 
(achieved by setting $m=\ell+1$ and $s=1$ in the sum),
remembering that the overlap in the RR sector carries a
phase $e^{i\pi Q}=e^{i{\pi\over h}(\ell+1)}$, 
and moreover suppressing the $S$ labels on the l.h.s,
\interdef\ becomes
$$
\big(I^{ADE}_{{L_1},{L_2}}\big)_{M_1}^{\ M_2}=
%(-1)^{(S_2-S_1)/2}\!\!\!
\sum_{\ell+1\in\cE(G)}
{
(\psi_{L_1}^{\ (\ell)})^*\psi_{L_2}^{\ (\ell)}
\over
{{S_{(0,0,0)}^{\ \ (\ell,\ell+1,1)}}}
}\,
e^{i{\pi\over h}(\ell+1)(M_2-M_1+1)}
e^{-i{\pi\over 2}(S_2-S_1)}
\ .
\eqn\ISSpp
$$
The exponential involving the $M_i$ can be rewritten in terms of a
sine function, so that we get: 
$$
\eqalign{
\big(I^{ADE}_{{L_1},{L_2}}\big)_{M_1}^{\ M_2}\ &=\
(-1)^{(S_2-S_1)/2}\!\!\!\sum_{\ell+1\in\cE(G)}
(\psi_{L_1}^{\ (\ell)})^*\psi_{L_2}^{\ (\ell)}
S^{\ \ell}_{M_2-M_1+1} \big(S_0^{\ \ell}\big)^{-1}
\cr
&\equiv\ (-1)^{(S_2-S_1)/2} \,
N_{{L_1},{L_2}}^{\ \ M_2-M_1}\ ,
}\eqn\ItoFus
$$
where  
$S_{\ell}^{\ (\ell')}\!=\! {1\over\sqrt2h}\sin\big[{\pi\over
 h}(\ell+1)(\ell'+1)\big]$
are the $S$-matrices and $N_{{L_1},{L_2}}^{\ \ell}$ 
nothing but the boundary fusion coefficients of
$\widehat{SU(2)}_{h-2}$ for the corresponding $ADE$ type modular
invariant \BPPZ. 
The fact that a computation involving an $\N=2$  superconformal
minimal model yields the fusion coefficients of an $SU(2)$ WZW
model, is perhaps not too surprising in view of the observation \OV\
that the $\N=2$ minimal model, when tensored with the non-compact
$SL(2)$ theory, turns into an $SU(2)$ WZW model plus some additional
free fields.

We are not yet done, because we still need
to identify the group theoretical meaning of the
intersection index $I^{ADE}_{L_1,L_2}$. For this we can make
use of certain group theoretical facts \Kos\ that involve the Coxeter
element $w$ of the Weyl group, $\cW(G)$. They have been
very useful in the past in the context of integrable systems 
\refs{\LW,\dorAB,\dorC},
and indeed we will draw on some of the results of these papers.

Recall that the Coxeter element is the unique (up to conjugation)
element of the Weyl group that is of order $h=h(G)$ (that's why $h$ is
called the Coxeter number).  Physically, $\ZZ_h$ is simply the
$R$-symmetry of the LG potentials \LGpotentials. If we denote by $r_i$
a Weyl reflection in the simple root $\a_i$, then a useful
representation of the Coxeter generator is given by $w=w_{(\bullet)}\cdot
w_{(\circ)}\in \ZZ_h\subset\cW(G)$, where
$$
w_{(\bullet)}\ =\ 
\prod r_\bullet\ ,\qquad\ 
w_{(\circ)}\ =\ 
\prod r_\circ\ .
\eqn\coxdef
$$
The important point for us is that the roots of $G$ decompose into
$r={\rm rank}(G)$ orbits of the Coxeter element, which are one-to-one
to the nodes of the Dynkin diagram $\Delta(G)$. More specifically, one
can identify a unique representative $\phi_i$, $i=1,...,r$ for each
orbit, by demanding that if $\phi_i$ is a positive root,  then
$w\cdot\phi_i$ is a negative root. Concretely, one can write the
following explicit representation \dorC:
$$
\phi_\bullet\ =\ w_{(\circ)}\cdot\a_\bullet\ ,\qquad
\phi_\circ\ =\ \a_\circ\ . 
\eqn\orbitreps
$$
In this way we have a direct correspondence
between the boundary state labels $L_i$
and the $\ZZ_h$ orbits of the roots of $G$.

We are now ready to make use of the following
formula that was proven in ref.\ \dorC:
$$
N_{{L_1},{L_2}}^{\ \ell+u_{1}-u_{2}}\
=\
\langle\lambda_{L_1},w^{-\ell/2}\phi_{L_2}\rangle\ ,
\eqn\thankgadforthisformula
$$
where $\lambda_L$ is the fundamental weight dual to the simple root
$\a_L$, and $u_{i}=0$ iff $L_i\in \{L_\bullet\}$ and
$u_{i}=1$ iff $L_i\in \{L_\circ\}$. 
This formula makes the crucial step in translating the ADE
boundary fusion coefficients into inner products in  the ADE weight
space.

In the final step the non-compact piece $z^{-h}$ of the LG  tensor
products~\LGpotentials\ comes into play. As was argued in \WL, it
contributes a factor $(1-w^{-1})$ to the intersection index
and this can  be used to convert the fundamental weight to the
Coxeter orbit representative \Kos: $\phi_L=(1-w^{-1})\cdot\lambda_L$. We
thus obtain alltogether (suppressing the $S$ labels):
$$
\big(I^{ALE}_{{L_1},{L_2}}\big)_{M_1}^{\ M_2}\ =\ 
\langle\phi_{L_1},w^{(M_1-M_2+u_{2}-u_{1})/2}\phi_{L_2}\rangle
\ ,
\eqn\finaleq
$$
where $L_i=1,...,r$ and $M_i=-h+1,...,h$.  Note that the $u_{i}$
implement the ``grading'' induced by the bi-coloration of the Dynkin
diagram, which can be expressed in terms of the selection rule:
$u_i+M_i+S_i=0$ mod 2 (generalizing the selection rule
$L_i+M_i+S_i=0$ mod 2 for the $A$-series).
Since \finaleq\ is periodic in
the $M$ labels, we can rewrite it in terms
of the cyclic $2h\times 2h$ shift generator $\gamma(2h)$ 
(with $\gamma(2h)^{2h}=1$) as follows:
$$
I^{ALE}_{{L_1},{L_2}}\ =\ 
\sum_{k=0}^{h-1}\langle\phi_{L_1},w^k\phi_{L_2}\rangle\,
\gamma(2h)^{2k+u_{2}-u_{1}}\ ,
\eqn\gform
$$
which makes the $\ZZ_h$ symmetry of the potentials \LGpotentials\ manifest.

Eq.\ \gform\ is precisely what was expected: the boundary states
are one-to-one to the roots of $G$ (organized in orbits of the $\ZZ_h$
Coxeter symmetry labelled by $L_i$),  and moreover their intersection
index gives the inner product between root vectors. This shows that
the $\N=2$ minimal model boundary states indeed correspond to
$D$-branes wrapped around the primitive cycles of the ALE space.

%%%%%%%%%%%%%%%%%%%%%%%%%%%%%%%%%%%%%%%%%%%%%%%%%%%%%%%
\chapter{Affine extension including D0-branes}
%%%%%%%%%%%%%%%%%%%%%%%%%%%%%%%%%%%%%%%%%%%%%%%%%%%%%%%

Note that we have
obtained a finite spectrum of BPS states (wrapped $D2$ branes) on the
ALE space, in one-to-one correspondence to the finite number of diagonal 
primary fields of the ADE type $\N=2$ superconformal minimal models.
It is given in terms of the roots of the gauge group, precisely what is
expected for the rigid field theory limit of the type IIA string
compactification in which we send the string scale to infinity.
It is remarkable that the truncation of the CFT fusion rules 
has precisely the right structure to select within the root {\it lattice}
the finite root {\it system}.

However, at finite string scale we expect infinitely many further BPS
states, in particular $D0$ brane (or Kaluza-Klein) states with
arbitrary positive charge $n$. These are expected to extend the finite
root system to an affine one \naka\ 
(for a recent exposition, see e.g., \FM).
 In our framework they should arise from the coupling to the
non-compact sector of the LG models \LGpotentials. While at the moment
it is unclear to us how this works  for general groups, the situation
is much simpler for the $A$-series, where there is no distinction
between the bulk and boundary labels, $\ell$ and $L$.

That is, the sine functions in $\psi_L^{\ \ell}=S_L^{\ \ell}$ 
(c.f., eq.~\Smat)
allow for a natural periodic extension of the labels $L$ that appear in
the  fusion coefficients $N$ and intersection matrices $I$.  This
allows to formally generate an infinite spectrum of BPS charges,
starting from a basic set of ``fractional brane'' charges, $\vec
q_{(0)}$. More precisely, by choosing a different Coxeter element than
before (namely $w=\prod r_i$), we can group all the simple  roots in
the $\ell=0$ orbit, together with the extending root  $\a_h$: $\vec
q_{(0)}=\{\a_1,\dots,\a_{h-1},\a_h\}$;   this was the choice that was
made in ref.~\WL. We can then generate further charges by acting with
the step generator \DR
$$
t_{\ell}\ =\ \sum_{k=-\ell/2}^{\ell/2}\gamma(2h)^{2k}
\eqn\tstep
$$
in the following way:
$$
\vec q_{(\ell)}\ = \vec q_{(0)}\cdot t_{\ell}\ .
\eqn\taction
$$
As shown in \WL, for the standard range $\ell=0,...,(h-2)$, this 
reproduces all the roots of $A_{h-1}$, with 
$I_{\ell_1,\ell_2}=t_{\ell_1}\cdot\cC\cdot {t_{\ell_2}}^t$ as their
intersection matrix ($\cC$ denotes the extended Cartan matrix
of $A_{h-1}$).

Extending now the range to arbitrary $\ell\in\ZZ_+$,
we first of all note  that there are gaps in the spectrum 
whenever $\ell=h-1$ (mod $h$),
which is where the intersection index vanishes: 
$I_{h-1({\rm mod}\ h),\ell_2}=0$. Moreover we find that (up to ordering
of the components): $\vec q_{(\ell+nh)}=\vec q_{(\ell)}+n\delta$,
$n\in\ZZ_+$, where $\delta\equiv \sum_{i=1}^h \a_i$
(this follows from the fact that the highest root $\delta$ is 
associated with the corner entry of the shift generator).
Accordingly, $I_{\ell_1+n_1h,\ell_2+n_2h}=I_{\ell_1,\ell_2}$.
This gives the requisite extension of the
root system to the one of $\widehat A_{h-1}$, obtained by adding
to $\{\a_i\}$ the imaginary simple root $\delta$, with $\langle\delta,\delta\rangle=0$.
More precisely, the BPS spectrum we get in this way corresponds
to the positive roots of $\widehat A_{h-1}$:
$$
\big\{\hat\a_+\big\}\ =\ \
\big\{\a_++n\delta, n\geq0\big\}
\cup
\big\{-\a_++n\delta, n>0\big\}
\cup
\big\{n\delta, n>0\big\}\ ,
\eqn\posaffine
$$
where $\a_+$ are the positive roots of $A_{h-1}$. Physically, the first
set corresponds to the wrapped $D2$-branes with $\ell=0,...,[k/2]$
($k\equiv h-2$) plus their $KK$ excitations; the second set to the
branes with $\ell=[k/2]+1,...,k$, which may be viewed as
anti-$D2$-branes (associated with negative $\ell$ shifted by $h$) plus
$KK$ excitations. The last set corresponds of course to the pure $KK$
modes, or $D0$ bound states.

To motivate the above construction, note that the same algebraic
structure (a finite spectrum repeated infinitely many times with a null
state in between the copies) is familiar from 2d topological minimal
models coupled to  topological gravity. In fact it was argued in \OV\
that the LG models \LGpotentials\ are closely related to such systems.

More specifically, it is known for $A_{h-1}$ that the ground ring
\ground\ is given by the finite chiral primary ring of the (twisted)
$\N=2$ minimal model, times a tower of infinitely many gravitational
descendants \gravdesc: $\{x^\ell y^n\}$, $\ell=0,..,h-2$, and $n\in
\ZZ_+$. Actually, the extended spectrum can be entirely constructed from
within the LG model (so that $\{x^\ell y^n\}\cong\{x^{\ell+nh}\}$), by
suitably defining physical states in terms of equivariant cohomology
\equivc. This is the same structure  that we seem to find for the
boundary states, with the understanding that
each ring element leads to a whole $\ZZ_h$ orbit
of BPS states. 

Our point of view is, therefore, 
that the infinitely many BPS
states ($D2$-branes wrapped around the cycles of the ALE space plus $n$
$D0$-branes on top of that) should be nothing but the boundary
counterpart of the gravitationally extended chiral ring of the bulk.
This observation may be useful for finding a BCFT or boundary LG
description of the $D_0$ branes on $K3$.

%%%%%%%%%%%%%%%%%%%%%%%%%%%%%%%%%%%%%%%%%%%%%%%%%%%%%%%
\chapter{ALE fibrations and $\N=2$ Yang-Mills theories}
%%%%%%%%%%%%%%%%%%%%%%%%%%%%%%%%%%%%%%%%%%%%%%%%%%%%%%%

It is straightforward to extend our
results to $\N=2$ $d=4$ supersymmetric gauge theories of general ADE type,
and determine their non-perturbative BPS spectra at the origin of the
respective moduli spaces. As mentioned in ref.\ \WL, the corresponding
LG potentials are obtained by a particular fibration of the ALE spaces
over $\IP^1$, which amounts to setting $z^{-h}\to z_1^{-2h}+z_2^{-2h}$
in \LGpotentials. The intersection index of
$D3$-brane boundary states wrapping the compact 
cycles of the fibered ALE is obtained by including an 
extra factor of $(1-w^{-1})$ in \gform, and looks explicitly:
$$
I^{SW}_{{L_1},{L_2}}(G_h=ADE,N_f=0)\ =\ (1-g^{-1})^2
\sum_{k=0}^{h-1}\langle\lambda_{L_1},w^k\phi_{L_2}\rangle\,
g^{2k+u_{2}-u_{1}}\ ,
$$
with $g=\gamma(4h)^2$.
This is now identified with the intersection form of the vanishing cycles 
of the Seiberg-Witten curves, from which the strong coupling spectrum of dyons 
at the $\ZZ_{2h}$ symmetric point of the moduli space can be extracted, 
as explained in \WL.

The fibration procedure is not unique, and other fibrations of the same
ALE space give rise to $\N=2$ gauge theories with matter fields
\geomeng.\foot{For related work see also \pelc.} 
While most of the possibilities do not lead to tensor product
models and thus are hard to deal with, we find that $SU(N_c)$ with
$N_f=N_c-1$ massless matter multiplets is very simply described 
by:
$$
W_{N_c,N_f=N_c-1}^{SW}\ =\  {1\over {z_1}^{N_c+1}}+
{1\over {z_2}^{N_c(N_c+1)}}+x^{N_c}\ ,
$$
whose intersection form is: 
$$
I^{SW}_{{L_1},{L_2}}(N_c,N_f = N_c-1)\ =\ 
t_{L_1} (1 - g^{-N_c}) (1 - g^{-1}) (1 - g^{N_c+1}) t^T_{L_2}
$$
with $g = \gamma(2N_c(N_c+1))^2$.
Our boundary CFT methods allow a straightforward determination of the 
strong coupling dyon spectrum of these gauge theories
at the $\ZZ_{N_c(N_c+1)}$ symmetric origin of their
moduli spaces.

\goodbreak 
%%%%%%%%%%%%%%%%%%%%%%%%%%%%%%%%%%%%%%%%%%%%%%%%%%%%%%%
\ack
We like to thank Peter Mayr for discussions, and Cumrun Vafa
as well as the referee 
for comments on the manuscript.

%%%%%%%%%%%%%%%%%%%%%%%%%%%%%%%%%%%%%%%%%%%%%%%%%%%%%%%
\nobreak

%\bigskip
%\goodbreak
\refout
\end